\documentstyle[12pt,epsfig]{article}

\newcommand{\beq}{\begin{equation}}
\newcommand{\eeq}{\end{equation}}
\newcommand{\bea}{\begin{eqnarray}}
\newcommand{\eea}{\end{eqnarray}}
\newcommand{\nn}{\nonumber}

\newcommand{\gtrsim}{\ \rlap{\raise 
2pt\hbox{$>$}}{\lower 2pt \hbox{$\sim$}}\ }
\newcommand{\lessim}{\ \rlap{\raise 
2pt\hbox{$<$}}{\lower 2pt \hbox{$\sim$}}\ }

\newcommand{\pl}[1]{Phys. Lett. {\bf #1}}
\newcommand{\pr}[1]{Phys. Rev. {\bf #1}}
\newcommand{\prl}[1]{Phys. Rev. Lett. {\bf #1}}
\newcommand{\zp}[1]{Z. Phys. {\bf #1}}

\newcommand{\ijmp}[1]{Int. Jour. Mod. Phys. {\bf #1}}

\newcommand{\epj}[1]{Eur. Phys. J.{\bf #1}}
\newcommand{\nim}[1]{Nucl. Instr. Methods.{\bf #1}}
\makeatletter
\if@twoside
\else \oddsidemargin 00pt \evensidemargin 00pt
\fi
\let\@eqnsel = \hfil

\expandafter\ifx\csname 
mathrm\endcsname\relax\def\mathrm#1{{\rm #1}}\fi
\@ifundefined{mathrm}{\def\mathrm#1{{\rm #1}}}{\relax}

\makeatother

\begin{document}

\title{\vskip-2.5truecm{\hfill \baselineskip 14pt {{
\small  \\
\hfill MZ-TH/98-52 \\ 
\hfill December  1998}}\vskip .9truecm}
 {\bf Neutrino Mixing and Masses from \\Long Baseline and Atmospheric \\
Oscillation Experiments}}

\vspace{5cm}

\author{Gabriela Barenboim\footnote{\tt 
gabriela@thep.physik.uni-mainz.de} 
\phantom{.}and Florian Scheck\footnote{\tt 
Scheck@dipmza.physik.uni-mainz.de}
 \\  \  \\
{\it  Institut f\H ur Physik - Theoretische 
Elementarteilchenphysik }\\
{\it Johannes Gutenberg-Universit\H at, D-55099 Mainz, 
Germany}
\\
}

\date{}
\maketitle
\vfill

\begin{abstract}
\baselineskip 20pt
We argue that regardless of the outcome of future Long Baseline
experiments, additional information will be needed to unambiguously
decide among the different scenarios of neutrino mixing.

We use, for this purpose, a simple test of underground
data: an asymmetry between downward and upward going events.
Such an asymmetry, in  which matter effects can be crucial,
tests electron and muon neutrino data separately and can be
compared
with the theoretical prediction without relying on any simulation
program.
\end{abstract}
\vspace{1cm}
\vfill
\thispagestyle{empty}

\newpage
\pagestyle{plain}
\setcounter{page}{1}

If the experimental evidence for neutrino oscillations and for
nonvanishing neutrino masses consolidates further, the primary goals of 
leptonic electroweak physics will be threefold: definite identification
of the flavour channels into which given initial states
oscillate, determination of the actual number and nature of neutral
leptonic states which participate in the mixing, and measurement of the 
(squared) mass differences of the states involved in the
mixing. Obviously, the analysis of any single experiment, or subset of 
experiments, in terms of only \emph{two} families can yield no more
than a parametrization of that experiment, or group of data, and no
special physical significance can be attached to the numerical
values of the parameters. In particular, the differences of squared
masses extracted in this way may lead to erroneous conclusions when
applied to the analysis of other experiments. Although there is
growing evidence, by now more or less accepted by the community, that
a scenario of three flavours $(e,\mu ,\tau )$ all of which mix
strongly, is compatible with all data showing neutrino anomalies, a 
mixing pattern involving a fourth, sterile, neutrino $\nu_S$ cannot
yet be excluded.

In the endeavour to answer these questions much hope is placed in the
planned or forthcoming long baseline (LB) oscillation
experiments \cite{lbl}. Although these experiments will be very important,
because of their much increased sensitivity as compared to short 
baseline oscillations, they are not sufficient to fix the parameters
of the mixed neutrino sector. As we show in this note, LB experiments
need complementary information for an unambiguous determination of
oscillation channels and mass differences and for discriminating
between the setup with three flavours and an alternative which
includes a sterile neutrino state. We point out that the up-down
asymmetry (zenith-antipode asymmetry) of atmospheric neutrino beams
might prove to be the complementary information that is needed,
provided the matter effects on the neutrinos coming from the antipode
are properly taken into account. We calculate these effects by
numerical integration of the neutrino's evolution equation and,
depending on the assumed scenario and on the neutrino energy,
find them to be important. As the mass distribution in the earth's
interior is well known, and as asymmetries are independent, to a large 
extent, of systematic errors, a clean analysis should be possible.

The interpretation of the modulation of neutrino events with zenith 
angle reported by Super Kamiokande in terms of 
$\nu_\mu - \nu_\tau$ oscillations,
obviously, would receive support by observing appearance of $\nu_\tau$ 
in LB experiments. This could be established either directly, through
observation of $\tau$-leptons, or indirectly, via an enhancement of the
neutral to charged current ratio, provided no large $\nu_e$ appearance
is found. The dependence on zenith angle of the
atmospheric neutrinos is confirmed by recent data from 
Super~Kamiokande. The analysis is performed separately on sub-GeV
($p$ $<$ 1.3 GeV) and multi-GeV ($p$ $>$ 1.3 GeV) data samples.
Electron-like ($\nu_e$ scattering) candidates and muon-like
($\nu_\mu$ scattering) candidates are presented separately
as a function of the zenith angle (see Fig. 4 of \cite{sk}). Looking at the
electron-flavour data, one notes that there appears to be a small
excess in the lowest $\cos\theta_z$ bin. This fact allows for the Super~Kamiokande
data to accommodate some $\nu_\mu \longrightarrow \nu_e$ at low 
$\Delta m^2$ which should be confirmed or disproved with further and more
precise data by Super~Kamiokande itself, or by the forthcoming LB
neutrino experiments.

With three generations of neutrinos, one expects a more complicated
oscillation phenomenology that includes transitions between all three
pairs of flavour channels. In particular, an experiment such as
Super~Kamiokande which measures $\nu_\mu$ disappearance, might in fact 
be seeing a combination of both $\nu_\mu \to \nu_\tau$ and
$\nu_\mu \to \nu_e$. In turn, if one analyzes the data in
terms of a \emph{two}-flavour scenario, say 
$\nu_\mu \to \nu_\tau$ to name the most popular,
one will extract a mixing angle and a unique, effective $\Delta {\cal {M}}^2$ 
which is some convolution of the two physical differences $\Delta M^2
= m_3^2-m_2^2$ and $\Delta m^2= m_2^2-m_1^2$ but by itself has no real
physical meaning.

Regarding LB experiments, and keeping in mind the lesson we have
learned from LSND \cite{lsnd}, we already know that even if there is a
large $\nu_\mu - \nu_\tau$ appearance signal it is not granted that
the energy distribution of the appearance signal alone would allow to
extract $\Delta m^2$ or, at least, to reduce significantly the
available parameter space. To witness, one should keep in mind that
even though LSND reports an excess of 50 events (plus a small
background) they are not able to distinguish the low and high $\Delta
m^2$ cases from their $\nu_\mu - \nu_e$ data.

It follows that a positive signal at any of the LB experiments 
cannot be taken as support for atmospheric $\nu_\mu$ oscillating
into $\nu_\tau$ unless the corresponding $\Delta M^2$ can be inferred
from  the \emph{same} signal. The reason for this is clear: The same
signal (for both appearance and disappearance) can be hiding a maximal 
$\nu_\mu - \nu_S$ oscillation with a small $\nu_\mu - \nu_\tau$
contamination (with $\sin^2(2\theta) \approx 10^{-3}$) or,
alternatively, a three neutrino mixing scheme with a low $\nu_\mu - \nu_e$
mixing rate. It is also important to notice that sterile neutrinos can only be 
invoked in cases where a deficit is observed, as opposed to a signal. 
So, sterile neutrinos may provide the explanation for the
deficits in atmospheric or solar neutrinos, but LSND must indeed be
$\nu_\mu - \nu_e$. 
At this point it may worth stressing that although the LSND signal
is somehow under suspicion because it lies very close to the region
already excluded by KARMEN \cite{kar}, 
one has to keep in mind that in their running
to date, KARMEN sees no event indicative for $\nu_\mu - \nu_e$ 
from the expected 3 background + 1 signal events (taking LSND
at face value). From this result the authors conclude that
not seeing any event allows them to exclude the LSND signal
at 90 \% C.L. . However, had they observed the expected background events, 
this sensitivity would not have been sufficient to exclude the LSND result 
to this confidence level.

Thus, independently of the outcome of LB neutrino experiments
we are urged to look for complementary information if an 
unambiguous interpretation of their future results is to be possible.
The complementary experiment we need must be one that allows us to
distinguish in a clear and unambiguous manner three possibilities
in interpreting the anomaly in atmospheric neutrino fluxes:
(i) the two-flavour interpretation in terms of the $\nu_\mu$
and $\nu_\tau$ channels, (ii) the hypothesis of the $\nu_\mu$ channel
oscillating into an otherwise sterile neutrino $\nu_S$,  and (iii) the 
scenario of three strongly mixing flavours with two differences of
squared masses $\Delta M^2 = m_3^2-m_2^2$ and $\Delta m^2=
m_2^2-m_1^2$ . In the absence of such a cross check, none of these
hypotheses can be excluded or taken for granted.

We argue, following a proposal by  Flanagan, Learned and Pakvasa
\cite{as}, that a quantity as simple as a directional asymmetry of
atmospheric neutrino fluxes might prove to be just what we need for
the purpose of distinguishing the alternatives described above. We
feel encouraged to explore this possibility by the recent sizeable
upgrading of Super~Kamiokande which already yields better statistics
than the one achieved in the previous, entire Kamiokande project.

The up-down asymmetry in $e$- or $\mu$-flavour events, $A^{(e)}$ or
$A^{(\mu )}$, respectively, is defined as
\bea
A^{(f)}= \frac{D^{(f)}-U^{(f)}}{D^{(f)}+U^{(f)}}\; ,\quad f=\mu ,e\, ,
\eea
where $D$ is the number of downward going events, produced in the
atmosphere in the zenith, for either electron or muon neutrinos,
while $U$ is the number of upward going events, stemming from the
antipode and having passed through the center of the earth. For the
case of $\mu$-flavour, $D$ and $U$ are
\bea
D^{(\mu )}(U^{(\mu )})= N_\mu^0 \left[ P_{\mu \mu}^{D(U)} + \frac{1}{r} 
P_{\mu e }^{D(U)} \right]
\eea
and similarly for electron flavour. Here  $N_{\mu (e)}^0$ is the
initial flux of muon (electron) neutrinos and 
$r=N_\mu^0/N_e^0$ is the expected ratio of fluxes without
oscillations. This ratio varies somewhat with energy 
and zenith angle \cite{concha} but typical values are $\sim 1/3$ for Multi-GeV
and $\sim 1/1.6$ for sub-GeV fluxes.

The important point to note is that matter effects on neutrino
oscillations due to the charged current interaction with electrons in
the earth can play a considerable role. As was shown long ago by Mikheyev
and Smirnov \cite{msw}, 
under certain conditions the presence of matter can lead
to a resonant amplification of the neutrino transitions, even if these 
transitions are strongly suppressed in vacuum.

Matter effects on the oscillation of neutrino beams which cross the
earth were studied previously in the two generation case
in \cite{two}, and in the three generation case in \cite{three}. Here
we study these effects for oscillations involving the three flavour neutrinos,
$\nu_e$, $\nu_\mu$ and $\nu_\tau$, by solving the neutrino evolution
equation. As the matter distribution of the earth neither is
homogeneous nor can be modeled by means of a simple analytic function,
we cannot rely on any simplifying assumptions and, therefore, we
obtain solutions of the evolution equation by numerical
integration. Note that with minor modifications our results
can be extended to oscillations of anti-neutrinos,
$\overline{\nu_e}$, $\overline{\nu_\mu}$ and $\overline{\nu_\tau}$.

In the presence of matter the neutrino wave function $\Psi^{(\nu
  )}=\{\mid\nu_e\rangle ,\mid\nu_\mu\rangle ,\mid\nu_\tau\rangle\}^T$
obeys the evolution equation
\bea
i \frac{d}{dl} \Psi^{(\nu )} = \frac{1}{2E} M(l) \Psi^{(\nu )}
\label{uno}
\eea
with
\bea
M (l)= U \pmatrix{m_1^2 &0&0 \cr 0 & m_2^2 &0 \cr
0&0&m_3^2 } U^\dagger + \pmatrix{a(l)&0&0\cr
0&0&0 \cr 0 &0&0 }
\eea
Here $m_i$ are the neutrino mass eigenvalues, and $U_{\alpha i}$
is the 3x3 mixing matrix in vacuum. The mixing matrix relates
the weak interaction states $\alpha$ and the mass eigenstates $i$ in
the leptonic sector, viz. 
\bea
\mid \nu_\alpha \rangle = \sum_{j=1}^3 U_{\alpha i} 
\mid n_i (p,m_i) \rangle  
\eea 
with $\mid n_i (p,m_i) \rangle$ denoting  the state vector of the mass
eigenstate with momentum $p$ and mass $m_i$ in vacuum.
It will prove to be convenient to use the following parametrization 
of the mixing matrix $U$ in vacuum,
\bea
U=\pmatrix{c_1c_3 & s_1c_3 & s_3 e^{-i \delta}\cr
       -s_1c_2 - c_1s_2s_3 e^{i \delta} & c_1c_2- s_1s_2s_3 e^{i \delta}
 & s_2c_3 \cr s_1s_2 - c_1c_2s_3 e^{i \delta} & - c_1s_2 - s_1c_2s_3 
e^{i \delta} &c_2c_3}
\eea

The background density of electrons, $N_e(l)$ induces a mass-like
interaction term for the  electron neutrino 
\bea
a(l)= 2 \, \sqrt{2} \, G_F \, E \, N_e(l)
\eea
with $E$ the neutrino energy.

For a concrete application to the case at stake, these equations
must be solved numerically. Note, however, that one can write down a
formal solution making use of the fact that the mass matrix in matter,
$M(l)$,
can be diagonalized at each position $l$ by means of a mixing matrix
$U_M (l)$. The formal solution reads
\bea
{\cal{A}} \left( n_i \left(l=0 \right) \rightarrow n_j \left(l=L
\right) \right) = P\; \exp \left[ - i \int_0^L dl \left(
\frac{1}{2 E}  \widetilde{M}(l) - i U_M^\dagger(l) \frac{d}{dl} U_M(l) \right)
\right]
\eea
where $ \widetilde{M}$ is a diagonal matrix containing the eigenvalues of $M(l)$
while $P$ denotes path ordering. Here
${\cal{A}} \left( n_i \left(l=0 \right)\rightarrow n_j \left(l=L
\right) \right)$ is the transition amplitude for the 
$i$th mass eigenstate created at $l=0$ to turn into the $j$th mass 
eigenstate detected at $l=L$. The probability of one flavour eigenstate
to propagate to another is then given by
\bea
{\cal{P}} \left( \nu_\alpha \left(l=0 \right)\rightarrow \nu_\beta 
\left(l=L\right) \right) = \vert \langle \nu_\beta \mid U_M(L) 
{\cal{A}} \left( n_i \left(l=0 \right)\rightarrow n_j \left(l=L
\right) \right) U_M^\dagger(0) \mid \nu_\alpha \rangle \vert^2
\label{dos} 
\eea
In the case of oscillations involving anti-neutrinos the amplitude
${\cal{A}}$ of finding the anti-neutrino $\beta$ at $l=L$ is obtained
from an evolution equation similar to Eq.(\ref{uno}).
The corresponding probabilities are obtained from Eq.(\ref{dos})
by making the formal change $U \rightarrow U^*$ and 
$a(l) \rightarrow - a(l)$. Note, however, that when there is a MSW
resonance in the particle sector due to level crossing, there is no
resonance in the antiparticle sector.

In our  analysis of the effects of earth on the neutrino beam from the 
antipode, we have used the density distribution as given by what is
called the preliminary reference earth model \cite{earth}, 
$\rho_E(r)$, $r$ denoting the distance from the center of the earth.
According to this model, $\rho_E(r)$ increases from an initial value of 
1.02 gr/cm$^3$ in the earth surface to its maximum of 13.1 gr/cm$^3$
in the center of the earth.  
There are basically nine regions in which $\rho_E(r)$ varies continuously.
The discontinuities of $\rho_E(r)$ at the borders of these regions are
described by step functions. The most pronounced of these jumps (all
of which are rather small) takes place at the border between the
mantle and the core where $\rho_E(r)$ changes by 4.3 gr/cm$^3$. This
happens at a distance $r$= 3480 km from the center of the earth.

In our calculation we assume that the ratio of the electron density to the
nucleon density is everywhere the same and is equal to $1/2$, $N_E = N_N/2$.
$N_N$ is the nucleon density and is given by $N_N = \rho_E N_A$ with
$N_A$ Avogadro's number.

The result of this tedious calculation can be summarized as follows:
For upgoing neutrinos originating from the antipode, matter effects 
affect substantially neutrino transitions if
\bea
10^3 \; \; \mbox{GeV}/\mbox{eV}^2 \leq   \;\;\;\; \frac{E}{\Delta m^2} \;\;
,\;\;\frac{E}{\Delta M^2}  \;\; \;\;\leq 10^{5} 
\; \; \mbox{GeV}/\mbox{eV}^2
\eea
or equivalently
\bea
10^{-14} \; \; \mbox{eV} \leq   \;\;\;\; \frac{\Delta m^2}{E} \;\;
,\;\;\frac{\Delta M^2}{E}  \;\; \;\;\leq 10^{-12} 
\; \; \mbox{eV}\; .
\eea
(As $\Delta m^2\ll\Delta M^2$ we have taken $ m_3^2 -m _1^2 \simeq \Delta M^2 $).
The magnitude of the interval (10) or (11) is determined by the
range of values of the electron number density that we find in earth
when crossing it along a diameter. Remember that matter effects are
important only if the interaction energy squared (7) 
\bea
 a(l) \approx
7.7 \cdot 10^{-5} \mbox{eV}^2 \; \left( \frac{\rho}{\mbox{gr/cm$^3$}} \right)
\; \left( \frac{E}{\mbox{GeV}} \right)
\eea
is comparable to or larger than either $\Delta m^2$ or $\Delta M^2$,
cf. Eqs.~(13) and (14) below. The enhancement typically shows up in
the dependence of the probability of a given transition on the
neutrino energy as an irregular sequence of two or three well 
pronounced local maxima with different heights.

For $E/\Delta m^2 (M^2)\gg$ 10$^{5} \mbox{GeV/eV$^2$}$  
or, equivalently, $\Delta m^2 (M^2)/E \ll$ 10$^{-14} \mbox{eV}$,   
the resonant densities 
$\vert m_i^2-m_j^2\vert\cos (2\theta )/(2\sqrt{2}G_FE)$ are much 
smaller than the electron density in the earth which, as mentioned
before, varies from 1 to 13 gr/cm$^3$. In this case the
charged-current interaction with electrons in the earth dominates
and oscillations are suppressed. 
For $E/\Delta m^2 (M^2) \ll$  10$^{3} \mbox{GeV/eV$^2$}$  or,
equivalently,  $\Delta m^2 (M^2)/E \gg $ 10$^{-12} \mbox{eV}$,
the resonant densities are much larger than the electron density and
neutrinos oscillate like in vacuum.   

When a beam of electron neutrinos crosses the earth there can potentially be two
resonances, in the $\nu_e\to\nu_\mu$ and the $\nu_e\to\nu_\tau$
channels. In the case of muon neutrinos there can only be one
resonance, the one in the $\nu_\mu\to\nu_e$ channel. In the
parametrization (6) the resonance densities are
given by
\bea
N_{\nu_e \to \nu_\mu}^{(res)} &= &
\frac{\Delta m^2 \cos 2\theta_1}{2 \sqrt{2} E G_F } = N_{\nu_\mu \to
  \nu_e}^{(res)}\, ,  \\
&& \nonumber \\
N_{\nu_e \to \nu_\tau}^{(res)} &= &
\frac{\Delta M^2 \cos 2\theta_3}{2 \sqrt{2} E G_F }\; .
\eea
These resonances would show up wonderfully if the vacuum mixing
angles, $\sin\theta_1 \ll 1 $ and $\sin\theta_3 \ll 1 $ were small,
\cite{small}, provided they are sufficiently separated \cite{sep},
i.e. provided
\bea
\Delta M^2 -\Delta m^2 \gg \left( \Delta M^2 \sin\theta_3 +
\Delta m^2 \sin\theta_2 \right)\; .
\nonumber
\eea

It can be shown that the resonances take place at densities at which 
the differences of local mass eigenvalues $M_2(l) - M_1(l)$ and
$M_3(l) - M_1(l)$ are minimal. One advantage of the parametrization
(6) we have chosen for the mixing matrix  
is that $M_1(l)$,  $M_2(l)$ and $M_3(l)$ do not depend on the CP
violating phase $\delta$ and therefore the very existence of
these resonances as well as the electron number densities at which
they occur are independent of $\delta$. Indeed, the eigenvalues of
$M(l)$ are given by
\bea
M_1^2(l) &=& m_1^2 + \frac{A}{3} - \frac{1}{3} \sqrt{A^2 -3B} \left(
\cos\omega + \sqrt{3} \sin\omega \right) \\
M_2^2(l) &=& m_1^2 + \frac{A}{3} - \frac{1}{3} \sqrt{A^2 -3B} \left(
\cos\omega - \sqrt{3} \sin\omega \right) \\
M_3^2(l) &=& m_1^2 + \frac{A}{3} +\frac{2}{3} \sqrt{A^2 -3B}
\cos\omega 
\eea
where
\bea
A &=& 2 \Delta m^2  + \Delta M^2 + a(l) \\
\nonumber \\
B &=& \left(\Delta M^2  + \Delta m^2 \right)\; 
\Delta m^2 + a(l) \left[ \left(\Delta M^2 + \Delta m^2\right)\; 
\cos^2 \theta_3 +  \right. \nn\\
&& \left.
 \Delta m^2 \left( \cos^2 \theta_3 \; \cos^2 \theta_1 + \sin^2 \theta_3
\right) \right] \\
\nonumber \\
C &=& a(l)\; \left(\Delta M^2 + \Delta m^2 \right)\; 
\Delta m^2 \; \cos^2 \theta_3 \; 
\cos^2 \theta_1 \\
\nonumber \\
\omega &=& \frac{1}{3} \arccos \frac{2A^3 - 9 A B + 27 C}
{2 \left( A^2 -3 B \right)^\frac{3}{2}\; .}
\eea
Note, however, that the transition probabilities do depend in a
nontrivial way on $\delta$.

We now turn to the analysis of the up-down asymmetry for both electron 
and muon neutrinos. We calculate this asymmetry for the three scenarios 
listed in the introduction, i.e. the two-flavour interpretation,
oscillation into a sterile $\nu_S$ channel, and strong mixing of three
flavours with two rather different mass differences, all of which would
give similar and hardly distinguishable signals in LB experiments.

The two-flavour mixing is the simplest possible case, although it is bound
to explain the atmospheric neutrino anomaly only, while discretely keeping 
silent about the solar neutrino deficit and the LSND result.
Here, the electron neutrino flux is unaffected and therefore, the corresponding
electron asymmetry $A^{(e)}$ is essentially zero. In order to calculate the muon 
asymmetry, we have to take the limit $\sin \theta_1 \rightarrow 0$  and
$\sin \theta_3 \rightarrow 0$  in the three neutrino case outlined
before. Although $A^{(\mu )}$ is nonzero, matter effects are
negligible in this case. 

In the scenario of three strongly mixing flavours \cite{nos} the
larger of the two differences $\Delta M^2 \equiv m_3^2-m_2^2= .3$ eV$^2$ 
is found to account for both, the atmospheric
anomaly at low energy as well as the observations of LSND. The second
mass difference $\Delta m^2 \equiv m_2^2-m_1^2\simeq 
10^{-3} - 10^{-4}$  eV$^2$ explains the solar neutrino
deficit but it is also called to play an important role in the atmospheric 
neutrino anomaly for upward going events. In this case, as of the two
would-be resonances the one corresponding to $\nu_e \to \nu_\mu$ is
reached in the core of the earth, \emph{matter effects are dominant}.

Finally, another way to accommodate all observations is to introduce 
at least one additional neutrino state $\nu_S$ which mixes with the
active neutrinos but otherwise is sterile. In Ref \cite{ste} the four
possible types of neutrino spectra were considered. It was shown there
that only two out of all possibilities  are compatible with the
existing bounds. In these two schemes, the four neutrino masses are
grouped in two pairs of closeby masses, the two groups being separated by a gap
of the order of 1 eV. Thus, it is assumed that $\Delta m_{41}^2=1$
eV$^2$ (we again use the abbreviation $\Delta m_{ij}^2 = m_i^2
-m_j^2$). In one scheme, say (A), the choices are $\Delta m^2_{21} \simeq
10^{-3}$ eV$^2$ (in view of explaining  the atmospheric neutrino
anomaly) and $\Delta m^2_{43} \simeq 10^{-10}$ or $10^{-5}$ eV$^2$
(needed for the suppression of the solar electron neutrinos). Here the
numbers refer to the vacuum oscillation solution and the MSW solution,
respectively.  In the other scheme, say (B), the roles of $\Delta m^2_{21}$
and $\Delta m^2_{43}$ are reversed. A detailed analysis of these
models as well as the bounds following from them can be found in
\cite{det}. 

What is important to notice is that the effective Hamiltonian of the
interaction of neutrinos with matter in the case of models containing
active and sterile neutrinos, has an additional neutral-current term,
apart from the usual charged-current term, which is proportional to the 
background density of neutrons, and therefore has half the
charged-current term strength. 

Cutting this long story short, in both models (A) and (B), oscillations
between the electron neutrino and the sterile neutrino are not affected
by matter effects. In the case of model (A) the $\nu_e \to\nu_\mu$
oscillations are in the appropriate range for resonant effects to
occur while crossing the earth and, therefore, electron neutrino beams
will oscillate and get depleted substantially. The resulting electron
asymmetry $A^{(e)}$ would be strikingly different from the three-flavour
case: its property of having the opposite sign in the two scenarios
would  make it a unique and easy criterion to distinguish these
alternatives. Furthermore, in these models with a sterile neutrino
$A^{(e)} $ and $A^{(\mu )}$ will differ only slightly from one
another; they are basically similar and always have the same sign.

Figure 1 shows our results for the electron neutrino asymmetry
$A^{(e)}$, Fig.~2 gives the corresponding results for the muon
neutrino asymmetry $A^{(\mu )}$, as a function of the neutrino energy
and for the three models described above. It is clear that given
sufficient statistics the three scenarios can be clearly distinguished
by both, the energy dependence and the relative signs of the
asymmetries $A^{(\mu )}$ and $A^{(e)}$. In particular, it is
noteworthy that in the case of the electron asymmetry, for which the
two-flavour oscillation model predicts a vanishing result, both
observation or non-observation of such an effect would be relevant.
Finally, although we have restricted our analysis to these three
models, it will be easy to calculate the expected asymmetries also for
other scenarios or for a different choice of parameters.

In conclusion, whatever the outcome of future Long Baseline neutrino 
experiments will be, complementary information will be crucial for an 
unambiguous interpretation of their results. A complemementary experiment 
is needed in order to distinguish in a clean manner between the 
$\nu_\mu \rightarrow \nu_\tau$, $\nu_\mu \rightarrow \nu_S$ and the 
three-flavour alternatives in interpreting the atmospheric neutrino anomaly.
For this purpose, we proposed to use the up-down asymmetries of electron and
muon neutrinos that are detected in underground detectors and that we
calculated by numerical integration of the evolution equation in
earth. These asymmetries have the virtue that they do not rely on any
simulations and that they require only identification of charged
particles and measurement of their energies and directions. 
Contrary to the zenith angle dependence, the asymmetries test electron
and muon neutrino data separately and do not require knowledge of the
fluxes. Given sufficient statistics, the asymmetries $A^{(e)}$ and
$A^{(\mu )}$, when added to the results of forthcoming long baseline
experiments, will help to distinguish the different scenarios that are 
being considered and will allow for an unambiguous determination of
the mixing matrix and the differences of squared neutrino masses.

\vspace{.5cm}

\begin{center}
{\bf Acknowledgements}
\end{center}

We are very grateful to S. Petcov for 
enlightening discussions. 
We also thank Ernst Otten and Christian
Weinheimer for discussions. A post-doctoral fellowship
of the Graduiertenkolleg ``Elementarteilchenphysik bei 
mittleren und hohen Energien"
of the University of Mainz is also acknowledged.

\newpage

\begin{figure}[!ht]
  \begin{center}
  \epsfig{file=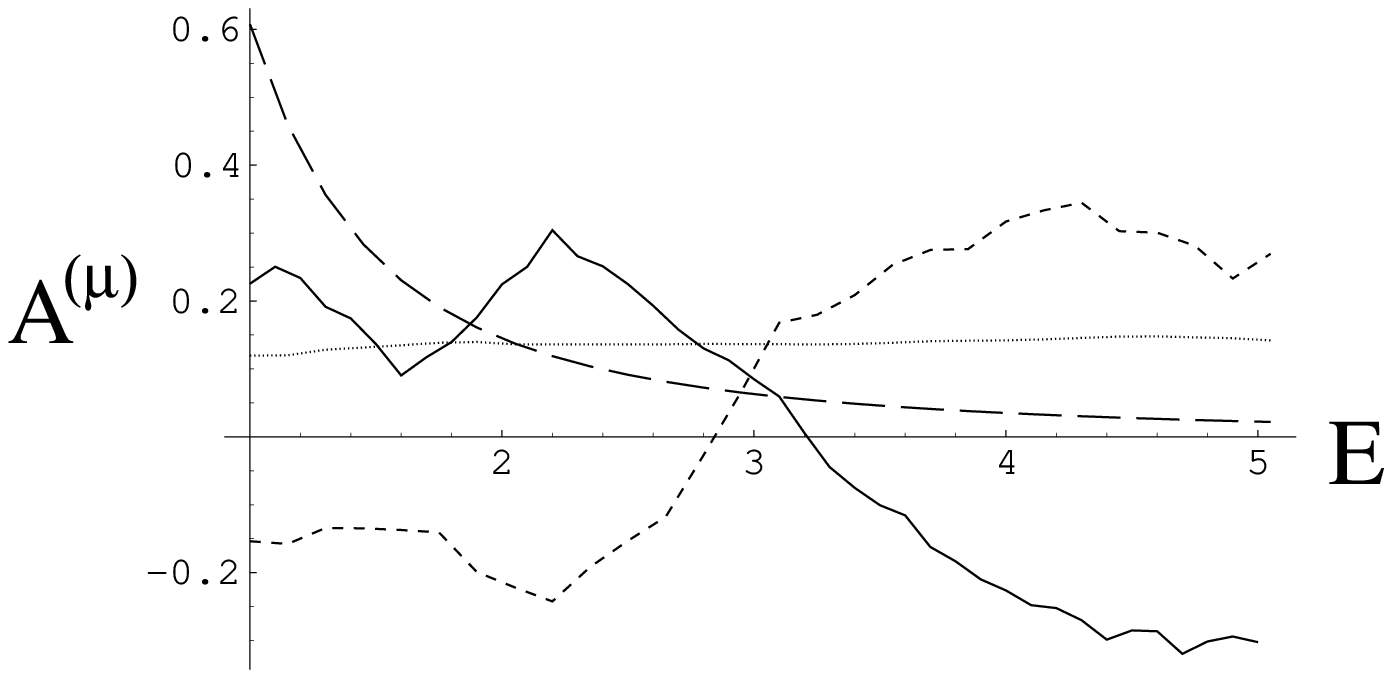,width=13cm}
\caption{Muon neutrino asymmetry, $A^{(\mu )}$, versus energy (GeV) for the two
flavour mixing scheme (long-dashed line), the three strongly mixed
flavours scenario (solid line) and the three active plus one sterile
neutrino model (A) (dotted line) and (B) (short-dashed line) } 
  \end{center}
\end{figure}
\vspace{1cm}
\begin{figure}[!ht]
  \begin{center}
  \epsfig{file=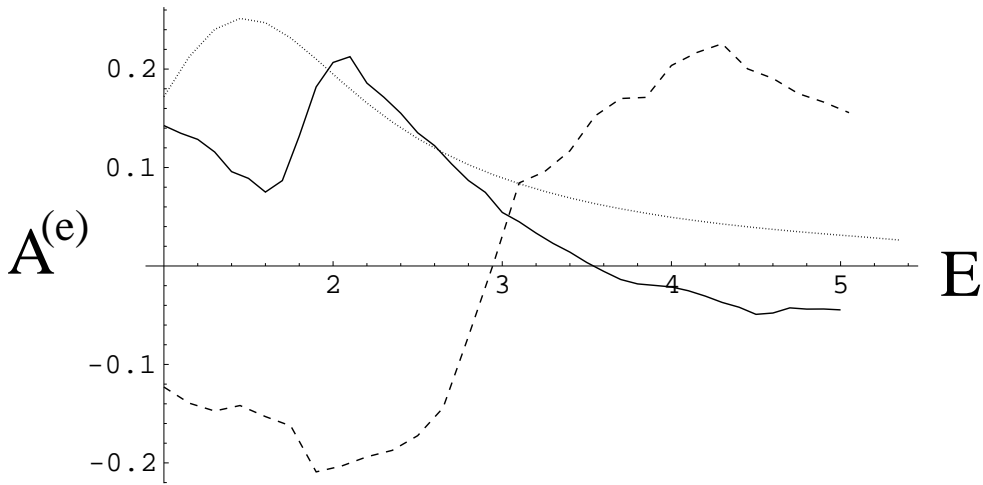,width=13cm}
  \parbox{15cm}{\caption{Electron neutrino asymmetry, $A^{(e)}$, versus 
energy (GeV) for the same models as in Fig. 1.}}
  \end{center}
\end{figure}

\end{document}